\documentclass[%
 aip,
 jmp,
 amsmath,amssymb,
 reprint,onecolumn
]{revtex4-1}

\usepackage{graphicx}
\usepackage{bm}

\usepackage[utf8]{inputenc}
\usepackage[T1]{fontenc}
\usepackage{mathptmx}
\usepackage{etoolbox}
\usepackage{mathrsfs}

\makeatletter
\def\@email#1#2{%
 \endgroup
 \patchcmd{\titleblock@produce}
  {\frontmatter@RRAPformat}
  {\frontmatter@RRAPformat{\produce@RRAP{*#1\href{mailto:#2}{#2}}}\frontmatter@RRAPformat}
  {}{}
}%
\makeatother

\begin{document}

\title{Interior spacetimes sourced by stationary differentially rotating irrotational cylindrical fluids: anisotropic pressure.}

\author{Marie-No\"elle C\'el\'erier}
\email{marie-noelle.celerier@obspm.fr}
\affiliation{Laboratoire Univers et Th\'eories, Observatoire de Paris, Universit\'e PSL, Universit\'e Paris Cit\'e, CNRS, 5 place Jules Janssen, F-92190 Meudon, France}

\date{5 July 2024}

\begin{abstract}
In a recent series of papers new exact analytical interior spacetimes sourced by stationary rigidly rotating cylinders of fluids  have been displayed. A fluid with an axially directed pressure has been first considered, then a perfect fluid, followed by a fluid with an azimuthally directed pressure, and, finally, by a fluid  where the pressure is radially oriented. The perfect fluid configuration has subsequently been extended to the case of differential rotation. In the present paper, three different cases of anisotropic pressure analogous to those studied for rigidly rotating motion are considered in turn for differentially rotating fluids. General methods for generating mathematical solutions to the field equations and physically well-behaved examples are displayed for the axial and azimuthal pressure cases. As regards radial pressure fluids, four classes of solutions naturally emerge from the corresponding Einstein's equations, among which one class, after being fully integrated,  exhibits physically well-behaved solutions.
\end{abstract}


\maketitle 


\section{Introduction} \label{intro}

To follow up on the exploration of exact analytical solutions of the field equations of general relativity (GR) defining the interior spacetimes of infinite cylinders of matter, we propose to consider here spacetimes sourced by stationary differentially rotating irrotational fluids with three differently directed anisotropic pressure. Indeed, in a first series of five papers, we have displayed a set of exact solutions, corresponding to rigidly rotating cylindrically symmetric sources with non vanishing rotation tensor. The perfect fluid case was analyzed as an extension of a previous work by Krasinski \cite{K74,K75a,K75b,K78} where, by using a completely different method, we have been able to fully integrate the field equations \cite{C23a}. Two other papers in our series were devoted to the study of fluids with axially directed pressure \cite{C21,C23b}. Azimuthal pressure was applied to the fluid considered in the following paper \cite{C23c}. Finally, the last rigidly rotating case involving a purely radial pressure was solved in the last paper of this first series \cite{C23d}.

Then, we turned to differentially rotating fluids. Indeed, the rotation of the astrophysical objects observed in the Universe is closer to differential than rigid. Since the number of extra degrees of freedom of the problem is increased as regards the rigidly rotating case, we proposed to use one of them by choosing to impose an irrotational motion to the fluid.  This choice has been implemented into the first paper of this new series \cite{C23e}, and is again made here, where it is discussed at length in Sec. \ref{feq}.

The use of cylindrical symmetry must also be considered as a relevant simplifying tool. This choice has been fully justified in the previous papers dealing with this issue. We merely want to recall here that, even though a rotating cylinder of matter, infinitely extended in its axis direction, and therefore generating a spacetime deprived of asymptotic flatness, can hardly represent as such a standard astrophysical object, the display of new analytically exact solutions can be essential to the understanding of fundamental gravitational features. Moreover, as it is always the case when one wants to consider any exact solution of GR as a representation of some corresponding physical process, the solutions described here might be used for the study of the nearly-cylindrical sources encountered in the Universe.

Moreover, it must be stressed that local anisotropy appears in a number of different physical configurations which have been described at length by, e. g., Herrera and Santos \cite{H97}. We can cite, for example, exotic phase transitions in the framework of highly dense systems or dissipative processes in some stellar evolutionary scenarios. As it has been shown by Herrera \cite{H20}, dissipative processes of the kind expected in stellar evolution always tend to produce pressure anisotropy, even if the system is evolving from an initial isotropic state. Hence the final state should, in principle, exhibit pressure anisotropy, since the specific isotropization process occurring in a collapse scenario is not relevant in the stationary rotation case considered here. Note that the generation of pressure anisotropy during the evolution concerns not only the spherically symmetric case, but also the axially symmetric case, hence the case of cylindrical symmetry.

The present paper displays the results obtained for the same three configurations previously considered for rigidly rotating fluid sources with an anisotropic pressure. General methods for generating new mathematical solutions to the field equations are proposed for the axial and the azimuthal cases and fully integrated classes with physically robust features are presented as exemplifying these recipes. Their mathematical and physical properties are analyzed. As regards fluids with radially directed pressure, four classes of solutions naturally emerge from the field equations. One of them is fully determined and its physical features analyzed and validated. Another is discarded owing to its inappropriate properties. The two others are reduced to a set of simplified differential equations which cannot be analytically integrated as such, but could be used in a numerical integration context.

The paper is organised as follows. In Sec.\ref{ge}, the equations displayed in previous papers and still of use here are recalled and specialized to the case of fluids with anisotropic pressure. A justification of the simplification issued from the irrotational assumption is also given. Section \ref{axial} is dedicated to a further specialization of the equations corresponding to a fluid with purely axial pressure. A general method for solving them is described. It is exemplified by the integration of a class of solutions, named class A. It is shown that this class of solutions is well-behaved for a large range of parameter values. In Sec.\ref{azimuthal}, the interior  spacetimes generated by a fluid with an azimuthally directed pressure are considered. A general method for integrating the field equations when the ratio $h$ of the magnitude of the pressure to that of the energy density depends on the radial coordinate is displayed and exemplified through a class denoted 1. Then, the case where this ratio is constant, named class 2, is fully integrated, while the special value $h=1$ is ruled out. Finally, the case of a fluid with radial pressure is considered in Sec.\ref{radial}. Four different classes of solutions are directly identified from the field equations. However, after analysis, only class I, exhibiting a constant negative pressure, possibly assimilated to a cosmological constant, is retained as including proper fully integrated GR spacetimes. Section \ref{concl} is devoted to the conclusions.

\section{Cylindrical differential rotation: generalities} \label{ge}

The general properties of interior spacetimes sourced by stationary rotating fluids have been established in \cite{CS20} and specialized to the case of differential rotation in \cite{C23e}. In particular, it has been shown that the problem is under-determined and that a number of degrees of freedom are left for a full specification of the issue. One of them has been employed such as to simplify the set of equations to be solved by making the rotation tensor of the fluid vanishing \cite{C23e}. It is also used here where irrotational fluids are again considered. The equations displayed in \cite{C23e} and being of use in the present article are recalled below and specialized to fluids with anisotropic pressure.

\subsection{Spacetime inside the source}

The stress-energy tensor of a stationary cylindrically symmetric anisotropic nondissipative fluid reads
\begin{equation}
T_{\alpha \beta} = (\rho + P_r) V_\alpha V_\beta + P_r g_{\alpha \beta} + (P_\phi - P_r) K_\alpha K_\beta + (P_z - P_r)S_\alpha S_\beta, \label{setens}
\end{equation}
where $\rho$ is the energy density of the fluid, $P_r$, $P_z$ and $P_\phi$ are the principal stresses. The four-velocity $V_\alpha$, and the spacelike four-vectors $K_\alpha$ and $S_\alpha$ satisfy
\begin{equation}
V^\alpha V_\alpha = -1, \quad K^\alpha K_\alpha = S^\alpha S_\alpha = 1, \quad V^\alpha K_\alpha = V^\alpha S_\alpha = K^\alpha S_\alpha =0. \label{fourvec}
\end{equation}
The bounding hypersurface $\Sigma$ being cylindrical, the spacelike Killing vector, $\partial_z$, is assumed to be hypersurface orthogonal, in view of a subsequent matching to an exterior Lewis metric. Hence, in geometric units, the line element can be written as
\begin{equation}
\textrm{d}s^2=-f \textrm{d}t^2 + 2 k \textrm{d}t \textrm{d}\phi +\textrm{e}^\mu (\textrm{d}r^2 +\textrm{d}z^2) + l \textrm{d}\phi^2, \label{metric}
\end{equation}
where $f$, $k$, $\mu$ and $l$ are real functions of the radial coordinate $r$ only. Owing to cylindrical symmetry, the coordinates conform to the following ranges:
\begin{equation}
-\infty < t < +\infty, \quad 0 \leq r, \quad -\infty \leq z \leq +\infty, \quad 0 \leq \phi \leq 2 \pi, \label{ranges}
\end{equation}
with the two limits of the $\phi$ coordinate topologically identified. These coordinates are denoted $x^0=t$, $x^1=r$, $x^2=z$ and $x^3=\phi$.

The four-velocity of the fluid, satisfying (\ref{fourvec}), is written as
\begin{equation}
V^\alpha = v \delta^\alpha_0 + \Omega  \delta^\alpha_3 , \label{4velocity}
\end{equation}
where the velocity, $v$, and the differential rotation parameter, $\Omega$, are functions of $r$ only. The timelike condition for $V^\alpha$ displayed in (\ref{fourvec}) becomes therefore
\begin{equation}
fv^2 - 2kv\Omega - l\Omega^2 -1= 0. \label{timelike}
\end{equation}
The two spacelike four-vectors defining the stress-energy tensor and verifying (\ref{fourvec}) are chosen as
\begin{equation}
K^\alpha = -\frac{1}{D}\left[(kv+l\Omega)\delta^\alpha_0 + (fv - k\Omega)\delta^\alpha_3\right], \label{kalpha}
\end{equation}
\begin{equation}
S^\alpha = \textrm{e}^{-\mu/2}\delta^\alpha_2, \label{salpha}
\end{equation}
with
\begin{equation}
D^2 \equiv fl + k^2. \label{D2}
\end{equation}

\subsection{Mathematical simplification and irrotational fluid} \label{feq}

By inserting (\ref{timelike})--(\ref{D2}) into (\ref{setens}), the components of the stress-energy tensor corresponding to the five nonzero components of the Einstein tensor are obtained. The five field equations for the inside $\Sigma$ spacetime follow \cite{CS20}. Together with (\ref{timelike}), one has thus six equations for ten unknown functions of $r$, namely, $f$, $k$, $\mu$, $l$, $v$, $\Omega$, $\rho$, $P_r$, $P_z$ and $P_\phi$. Therefore, four degrees of freedom are left. Four equations connecting the matter observables or assumptions on the metric functions would have thus to be imposed in order to solve the field equations. However, three among them can be partially integrated as follows \cite{CS20}:
\begin{equation}
\left(\frac{kf' - fk'}{D}\right)' = 2 \kappa (\rho + P_\phi) D \textrm{e}^\mu (k\Omega^2 - fv\Omega), \label{partint1}
\end{equation}
\begin{equation}
\left(\frac{lf' - fl'}{D}\right)' = 2 \kappa (\rho + P_\phi) D \textrm{e}^\mu (fv^2 + l\Omega^2),\label{partint2}
\end{equation}
\begin{equation}
\left(\frac{kl' - lk'}{D}\right)' = - 2 \kappa (\rho + P_\phi) D \textrm{e}^\mu (kv^2 + lv\Omega). \label{partint3}
\end{equation}
where the primes stand for differentiation with respect to $r$. An interesting simplification of the system would arise from setting one of the right hand side of the above equations to vanish. We choose as a possibility possessing a straightforward physically meaningful interpretation the vanishing of the right hand side of (\ref{partint3}) which gives
\begin{equation}
kv + l\Omega=0. \label{partint8}
\end{equation}
Inserting (\ref{partint8}) into (\ref{timelike}), one obtains
\begin{equation}
v^2 = \frac{l}{D^2}. \label{partint9}
\end{equation}
This is the choice that has been made for the study of the perfect fluid case \cite{C23e} and which is again made here. It amounts to using one among the four available degrees of freedom. Since (\ref{partint8}) yields a vanishing rotation tensor, as shown in Sec. \ref{genhydro}, the corresponding fluid is irrotational. From a mathematical point of view, this choice implies that (\ref{partint3}) can be integrated to give
\begin{equation}
kl' - lk' = 2c D, \label{partint11}
\end{equation}
where $c$ is an integration constant and where the factor 2 is retained for further convenience. Considered as a first-order ordinary differential equation in $k$, (\ref{partint11}) possesses, as a general solution,
\begin{equation}
k = l \left(c_k - 2c \int^r_{r_1} \frac{D(v)}{l^2(v) } \textrm{d} v \right), \label{partint12}
\end{equation}
where $r_1$ denotes an arbitrary value of the radial coordinate located between $r=0$ and its value at the boundary, $r_{\Sigma}$, and $c_k$ denotes a constant of integration.

Then, from (\ref{D2}) and (\ref{partint11}), we obtain
\begin{equation}
\frac{f'l' +k'^2}{2D^2} = \frac{l'D'}{lD} - \frac{l'^2}{2l^2} + \frac{2c^2}{l^2}. \label{partint14}
\end{equation}

\subsection{Conservation of the stress-energy tensor}

The conservation of the stress-energy tensor is implemented by the Bianchi identity. With $V^\alpha$ given by (\ref{4velocity}), and the space-like vectors $K^\alpha$ and  $S^\alpha$ given by (\ref{kalpha}) and (\ref{salpha}), respectively, this identity reduces to \cite{CS20}
\begin{equation}
T^\beta_{1;\beta} = P'_r - (\rho + P_\phi) \Psi + (P_r - P_\phi)\frac{D'}{D} + \frac{1}{2}(P_r - P_z)\mu'  = 0, \label{Bianchi1}
\end{equation}
with
\begin{equation}
\Psi = fvv' - k(v\Omega)' - l\Omega\Omega' = -\frac{1}{2} (v^2f' - 2v\Omega k' - \Omega^2l'), \label{psidef}
\end{equation}
which becomes, with the choice (\ref{partint8}),
\begin{equation}
\Psi = \frac{l'}{2l} - \frac{D'}{D}, \label{psi8}
\end{equation}
which, once inserted into the Bianchi identity (\ref{Bianchi1}), yields
\begin{equation}
 P'_r - (\rho + P_\phi)\left( \frac{l'}{2l} - \frac{D'}{D}\right) + (P_r - P_\phi)\frac{D'}{D} + \frac{1}{2}(P_r - P_z)\mu'  = 0. \label{Bianchi2}
\end{equation}

\subsection{Junction conditions} \label{junct}

Since the motion of the source is stationary, the Lewis metric \cite{L32} is well suited to represent the exterior spacetime, and the Weyl class of real metrics \cite{W19} is chosen such as to obtain junction conditions for a physically well-behaved fluid. These conditions have already been discussed at length for the metric (\ref{metric}) \cite{CS20,C21,D06}. Only their main resulting constraint acting on the interior spacetimes is of use here. 
 
Applying Darmois' junction conditions \cite{D27}, the radial coordinate of the pressure, $P_r$, must vanish on the boundary \cite{D06}. This is obvious for the the axial and azimuthal pressure cases where $P_r$ vanishes everywhere in spacetime, while it generates a constraint on the parameters of the solutions in the case of radial pressure.

\subsection{Hydrodynamical properties} \label{genhydro}

The hydrodynamical properties of rigidly rotating fluids have been analyzed and interesting features have been exhibited in \cite{C21,C23b,C23c,C23d}. For differential rotation, the assumption made in (\ref{partint8}) simplifies the expressions for the different hydrodynamical quantities. These expressions, valid for any configuration of the pressure, are displayed below.

The hydrodynamical vectors, tensors and scalars have been obtained \cite{CS20} with the use of the quantity $\Psi$ given by (\ref{psi8}). This quantity is used here in the determination of the only nonzero component of the acceleration vector that reads
\begin{equation}
\dot{V}_1 = -\Psi = \frac{D'}{D} -\frac{l'}{2l} . \label{hydr2}
\end{equation}
Hence the modulus of this acceleration vector comes easily as
\begin{equation}
\dot{V}^{\alpha}\dot{V}_{\alpha}= \textrm{e}^{-\mu} \left(\frac{D'}{D} -\frac{l'}{2l}\right)^2 . \label{hydr3}
\end{equation}

The generally nonzero components of the rotation or twist tensor, $\omega_{01}$ and $\omega_{13}$, have been derived in \cite{CS20}. Owing to (\ref{partint8}), both components vanish. Therefore, local rotation is null for differentially rotating fluids verifying (\ref{partint8}) and being thus irrotational.

The general forms of the nonzero components of the shear tensor have also been determined in \cite{CS20}. With the use of (\ref{partint8}), (\ref{partint11}) and (\ref{psi8}), and choosing a plus sign to determine the four-velocity $v$ from its squared expression (\ref{partint9}), they become
\begin{equation}
2 \sigma_{01} = \frac{2ck}{l\sqrt{l}}, \label{hydr9a}
\end{equation}
\begin{equation}
2 \sigma_{13} = \frac{2c}{\sqrt{l}}. \label{hydr10}
\end{equation}
By using (\ref{partint8}), (\ref{partint9}) and (\ref{partint11}), the general form of the shear scalar is therefore given by
\begin{equation}
\sigma^2 = c^2 \frac{\textrm{e}^{-\mu}}{l^2}. \label{hydr12}
\end{equation}

The above results confirm that, contrary to what happens in the rigid rotation case where the shear vanishes and the twist is nonzero, the shear is nonzero for differentially rotating irrotational fluids. While the fluid cylindrical shells rotate differentially, each element in the shell is in irrotational motion with respect to its neighbours.

\section{Specialization to a fluid with purely axial pressure} \label{axial}

The equation of state applying to an anisotropic fluid with axially directed pressure is
\begin{equation}
P_{\phi} = P_r = 0. \label{eosax}
\end{equation}

In this section, a general method for generating exact solutions to the field equations for an interior spacetime sourced by a differentially rotating irrotational fluid whose principal stresses $P_r$, $P_z$ and $P_\phi$ obey (\ref{eosax}) is displayed. This method is then exemplified by the determination of a class of physically relevant solutions. 

The choice of irrotational motion added to the double equation of state (\ref{eosax}) implies that one extra degree of freedom is still available for our purpose. Moreover, the general properties stated in Sec.\ref{ge} apply and the stress-energy tensor (\ref{setens}) becomes
\begin{equation}
T_{\alpha \beta} = \rho V_\alpha V_\beta + P_z S_\alpha S_\beta. \label{setensax}
\end{equation}

\subsection{Field equations} \label{fe}

By using (\ref{setensax}), we obtain the components of the stress-energy tensor corresponding to the five nonvanishing components of the Einstein tensor, and we can thus write the five field equations as 
\begin{equation}
G_{00} = \frac{\textrm{e}^{-\mu}}{2} \left[-f\mu'' - 2f\frac{D''}{D} + f'' - f'\frac{D'}{D} + \frac{3f(f'l' + k'^2)}{2D^2}\right]= \kappa \rho \frac{D^2}{l}, \label{G00ax}
\end{equation}
\begin{equation}
G_{03} =  \frac{\textrm{e}^{-\mu}}{2} \left[k\mu'' + 2 k \frac{D''}{D} -k'' + k'\frac{D'}{D} - \frac{3k(f'l' + k'^2)}{2D^2}\right] = 0, \label{G03ax}
\end{equation}
\begin{equation} 
G_{11} = \frac{\mu' D'}{2D} + \frac{f'l' + k'^2}{4D^2} = 0, \label{G11ax}
\end{equation}
\begin{equation}
G_{22} = \frac{D''}{D} -\frac{\mu' D'}{2D} - \frac{f'l' + k'^2}{4D^2} = \kappa P_z \textrm{e}^{\mu} , \label{G22ax}
\end{equation}
\begin{equation}
G_{33} =  \frac{\textrm{e}^{-\mu}}{2} \left[l\mu'' + 2l\frac{D''}{D} - l'' + l'\frac{D'}{D} - \frac{3l(f'l' + k'^2)}{2D^2}\right] =  0. \label{G33ax}
\end{equation}

\subsection{Bianchi identity}

Inserting the equation of state (\ref{eosax}) and the function $h(r)$ defined as $h(r)\equiv P_{z}(r)/\rho(r)$ into the Bianchi identity (\ref{Bianchi2}), we obtain
\begin{equation}
\frac{l'}{2l} - \frac{D'}{D} + \frac{h \mu'}{2} = 0. \label{Bianchi1ax}
\end{equation}

\subsection{General method for solving the field equations} \label{solvingax}

We begin with defining two new auxiliary functions $M(h(r))$ and $N(h(r))$ by
\begin{equation}
h \mu' = h' M =  \frac{N'}{N}, \label{MNdefax}
\end{equation}
which we insert into the Bianchi identity (\ref{Bianchi1ax}) and integrate as
\begin{equation}
l = c_D^2 \frac{D^2}{N}, \label{Bianchi.intax}
\end{equation}
where $c_D$ is an integration constant, squared for further purpose. 

By inserting (\ref{G11ax}), (\ref{Bianchi1ax}), (\ref{MNdefax}), and derivatives into (\ref{G33ax}), we obtain
\begin{equation}
\frac{D'}{D} = \frac{(1 + M h^2)h'}{3h(1+h)} - \frac{M'}{3M} - \frac{h''} {3h'}, \label{gen5ax}
\end{equation}
which we insert into (\ref{Bianchi1ax}) with (\ref{MNdefax}) implemented, which gives
\begin{equation}
\frac{l'}{l} = \frac{2(1 + M h^2)h'}{3h(1+h)} - \frac{2M'}{3M} - M h' - \frac{2h''}{3h'}. \label{gen6ax}
\end{equation}

Then, we define another auxiliary function $L(h(r))$ by
\begin{equation}
\frac{L'}{L} = \frac{(1 + M h^2)h'}{h(1+h)}, \label{gen7ax}
\end{equation}
which we insert into (\ref{gen6ax}), together with (\ref{MNdefax}) for $N$, which yields
\begin{equation}
\frac{l'}{l} = \frac{2L'}{3L} - \frac{2M'}{3M} - \frac{N'}{N} - \frac{2h''}{3h'}, \label{gen8ax}
\end{equation}
which can be integrated by
\begin{equation}
l = c_B^2  \frac{L^{\frac{2}{3}}}{M^{\frac{2}{3}} h'^{\frac{2}{3}}N}. \label{gen9ax}
\end{equation}
where $c_B$ is a constant of integration.

By inserting (\ref{gen9ax}) into (\ref{Bianchi.intax}), we obtain
\begin{equation}
D = c_B c_D  \frac{L^{\frac{1}{3}}}{M^{\frac{1}{3}} h'^{\frac{1}{3}}}. \label{gen10ax}
\end{equation}
Then, by inserting (\ref{gen9ax}) and (\ref{gen10ax}) into (\ref{Bianchi.intax}), we obtain $c_D=1$.

Now, (\ref{partint14}) and (\ref{Bianchi1ax}) substituted into (\ref{G11ax}) yield
\begin{equation}
\frac{h \mu'^2}{2}  + \frac{(1+h)}{2}\frac{l' \mu'}{l} + \frac{2c^2}{l^2} = 0, \label{gen12ax}
\end{equation}
where we substitute (\ref{MNdefax}), (\ref{gen6ax}), and (\ref{gen9ax}), which gives
\begin{equation}
h'^{\frac{1}{3}} = \frac{c_B^4}{6c^2} \frac{(1+h)}{h}   \frac{L^{\frac{4}{3}}}{M^{\frac{1}{3}} N^2} \left[\frac{h''}{h'} + \frac{1}{1+h} \left(\frac{h M}{2} - \frac{1}{h} \right) h' + \frac{M'}{M} \right]. \label{gen13ax}
\end{equation}

Then, we define a new function $g(h(r))$ through
\begin{equation}
h' = \frac{h^{\alpha} (1-h)^{\chi}}{(1+h)^{\beta}} g, \label{gen14ax}
\end{equation}
where $\alpha$, $\beta$, and $\chi$ are constants to be determined later. From (\ref{gen14ax}), we have
\begin{equation}
\frac{h''}{h'} = \frac{h^{\alpha} (1-h)^{\chi}}{(1+h)^{\beta}} \left( \frac{\alpha}{h} - \frac{\chi}{1-h} - \frac{\beta}{1+h} + \frac{\frac{\textrm{d}g}{\textrm{d}h}}{g}\right) g. \label{gen15ax}
\end{equation}
Then, we substitute (\ref{gen14ax}) and (\ref{gen15ax}) into (\ref{gen13ax}) and obtain
\begin{eqnarray}
g^{\frac{1}{3}} &=&  \frac{c_B^4}{6 c^2} \frac{h^{\frac{2 \alpha}{3} -1} (1-h)^{\frac{2\chi}{3}}}{(1+h)^{\frac{2 \beta}{3} - 1}}  \frac{L^{\frac{4}{3}}}{M^{\frac{1}{3}} N^2}\left\{ \left[ \frac{\alpha-1}{h} - \frac{\chi}{1-h}  \right. \right.\nonumber \\
&-& \left. \left.\frac{\beta-1}{1+h} + \frac{h M}{2(1+h)} + \frac{\frac{\textrm{d}M}{\textrm{d}h}}{M}\right] g + \frac{\textrm{d}g}{\textrm{d}h} \right\}. \label{gen16ax}
\end{eqnarray}
Here, we use the second degree of freedom, to make a choice allowing us to integrate easily the above equation, by setting to vanish the $g$ factor, inside the brackets, i. e.,
\begin{equation}
\frac{\alpha-1}{h} - \frac{\chi}{1-h} - \frac{\beta-1}{1+h} + \frac{h M}{2(1+h)} + \frac{\frac{\textrm{d}M}{\textrm{d}h}}{M} = 0. \label{gen17ax}
\end{equation}
This equation, considered as a first order differential equation in $M(h)$, possesses solutions of the form
\begin{equation}
M = \frac{a (1+h)^{\lambda}}{h^{\gamma} (1-h)^{\delta}}, \label{gen18ax}
\end{equation}
verifying
\begin{eqnarray}
2(\alpha -\gamma - 1) &+& 2(\delta + \lambda - \beta -\chi +1) h +2(\beta + \gamma + \delta - \alpha - \chi - \lambda) h^2 \nonumber \\
&+& a h^{2- \gamma} (1-h)^{1- \delta} (1+h)^{\lambda} = 0. \label{gen19ax}
\end{eqnarray}

Indeed, the problem of determining the set of parameters $\{ \alpha, \beta, \gamma, \delta, \chi, \lambda, a \}$ verifying (\ref{gen19ax}) is degenerate. However, once a given expression for (\ref{gen18ax}) is judiciously chosen, implying satisfactory expressions for $L$ and $N$, the other parameters follow from (\ref{gen19ax}), and (\ref{gen16ax}) becomes
\begin{equation}
\frac{\frac{\textrm{d}g}{\textrm{d}h}}{g^{\frac{1}{3}}} =  \frac{6 c^2}{c_B^4} \frac{(1+h)^{\frac{2 \beta}{3} - 1}}{h^{\frac{2 \alpha}{3} -1} (1-h)^{\frac{2\chi}{3}}}  \frac{M^{\frac{1}{3}} N^2}{L^{\frac{4}{3}}}, \label{gen22ax}
\end{equation}
which can be integrated so as to give $g^{\frac{2}{3}} (h)$ .

Once $g(h)$ is determined, $h'(h)$ follows from (\ref{gen14ax}). Therefore $r(h)$ is obtained by integrating the inverted $h'$ expression. Then $h'(h)$ is inserted into (\ref{gen9ax}) and (\ref{gen10ax}), together with $M(h)$, $L(h)$ and $N(h)$, so as to obtain $l(h)$ and $D(h)$ respectively. The metric function $\textrm{e}^{\mu}$ is obtained from the integration of the first equality in (\ref{MNdefax}). Then, $k(h)$ follows through (\ref{partint12}), and $f(h)$, by means of (\ref{D2}). By adding (\ref{G11ax}) and (\ref{G22ax}), one obtains
\begin{equation}
\frac{D''}{D} = \kappa P_z \textrm{e}^{\mu}, \label{gen24ax}
\end{equation}
which is used to obtain $P_z$. The energy density $\rho$ follows from the definition of the ratio $h$. The global rotation parameter $\Omega$ and the velocity $v$ proceed from (\ref{partint8}) and (\ref{partint9}).

We have thus obtained the full metric and the main characteristics of this set of spacetimes whose mathematical and physical properties can therefore be analyzed.

To exemplify this recipe, we consider now a particular realization of the $M$ function.

\subsection{Class A solutions} \label{A}

Most of the integrable solutions obtained for a given set of ${a, \lambda, \gamma, \delta}$ are irrelevant since they fail to exhibit proper physical features. The class A solutions described here appear however to be the simplest ones appropriate to this purpose. The degree of freedom is used to choose $a= -1$, $\lambda=\delta=0$, and $\gamma=2$. Therefore the $M$ function becomes
\begin{equation}
 M = - \frac{1}{h^2}. \label{C1}
\end{equation}

By implementing the method described in Sec. \ref{solvingax}, we obtain the fully integrated class A solutions under the form:
\begin{equation}
\textrm{e}^{\mu} = \textrm{e}^\frac{1}{2h^2}, \label{finC1}
\end{equation}
\begin{equation}
l = \sqrt{2} c \frac{(1+h)}{h \textrm{e}^{\frac{1}{h}}\left(\textrm{e}^{\frac{2}{h}} + c_g \right)}, \label{finC2}
\end{equation}
\begin{equation}
k = \sqrt{2} c  \frac{(1+h)}{h \textrm{e}^{\frac{1}{h}}\left(\textrm{e}^{\frac{2}{h}} + c_g \right)} \left(c_k - \frac{1}{2^{\frac{3}{4}}} \sqrt{\frac{c_N}{c}} \textrm{e}^{\frac{2}{h}} \right), \label{finC3}
\end{equation}
\begin{equation}
f = c_N \textrm{e}^{\frac{1}{h}} - \sqrt{2} c  \frac{(1+h)}{h \textrm{e}^{\frac{1}{h}}\left(\textrm{e}^{\frac{2}{h}} + c_g \right)} \left(c_k - \frac{1}{2^{\frac{3}{4}}} \sqrt{\frac{c_N}{c}} \textrm{e}^{\frac{2}{h}} \right)^2, \label{finC4}
\end{equation}
\begin{equation}
D = - 2^{\frac{1}{4}} \sqrt{c c_N}\sqrt{ \frac{(1+h)}{h \left(\textrm{e}^{\frac{2}{h}} + c_g \right)}}, \label{finC5}
\end{equation}
\begin{equation}
\rho = -\frac{h^2}{\kappa  \textrm{e}^\frac{1}{2h^2}(1+h)^5} \left[(2+2 h +h^2) \textrm{e}^{\frac{2}{h}} + c_g h^2 \right] \left(\textrm{e}^{\frac{2}{h}} + c_g \right)^2, \label{finC6}
\end{equation}
\begin{equation}
P_z = - \frac{h^3}{\kappa  \textrm{e}^\frac{1}{2h^2}(1+h)^5} \left[(2+2 h +h^2) \textrm{e}^{\frac{2}{h}} + c_g h^2 \right] \left(\textrm{e}^{\frac{2}{h}} + c_g \right)^2, \label{finC6}
\end{equation}
\begin{equation}
v = \frac{1}{\sqrt{c_N} \textrm{e}^{\frac{1}{2h}}}, \label{finC7}
\end{equation}
\begin{equation}
\Omega = - \frac{c_k}{\sqrt{c_N}\textrm{e}^{\frac{1}{2h}}} + \frac{\textrm{e}^{\frac{3}{2h}}}{2^{\frac{3}{4}} \sqrt{c}}, \label{finC8}
\end{equation}
\begin{equation}
h' = \frac{h^{\frac{7}{2}}}{(1+h)^{\frac{3}{2}}}\left(\textrm{e}^{\frac{2}{h}} + c_g \right)^{\frac{3}{2}}, \label{finC9}
\end{equation}
\begin{equation}
r-r_{\Sigma} = \int_{h_{\Sigma}}^h\frac{(1+u)^{\frac{3}{2}}}{ u^{\frac{7}{2}} \left(\textrm{e}^{\frac{2}{u}} + c_g \right)^{\frac{3}{2}}} \textrm{d}u , \label{finC10}
\end{equation}
\begin{equation}
\dot{V}^{\alpha}\dot{V}_{\alpha} = \frac{1}{4 \textrm{e}^{\frac{1}{2 h^2}}}\left[\frac{h}{(1+h)}\left(\textrm{e}^{\frac{2}{h}} + c_g \right) \right]^3, \label{finC11}
\end{equation}
\begin{equation}
\omega = 0, \label{finC12}
\end{equation}
\begin{equation}
\sigma^2 = \frac{h^2}{2(1+h)^2} \frac{\textrm{e}^{\frac{2}{h}}\left(\textrm{e}^{\frac{2}{h}} + c_g \right)^2}{\textrm{e}^{\frac{1}{2h^2}}}, \label{finC13}
\end{equation}
where $c_N$ and $c_g$ are constants of integration.

\subsubsection{Pressure and energy density} \label{peC}

The axisymmetry condition \cite{C23e}, applied to $l$ given by (\ref{finC2}), implies
\begin{equation}
h_0 = - 1 \qquad \textrm{or} \qquad h_0=0, \label{C20}
\end{equation}
where $h_0$ denotes the value of $h$ for $r=0$.

The energy density diverges on the axis if $h_0= -1$, while, for $h_0=0$, $\rho$ vanishes. Therefore, we must select $h_0 = 0$. In the vicinity of the axis, where $1+h>0$, up to $h_{\Sigma} > -1$, the weak energy condition is fulfilled provided that
\begin{equation}
c_g < - \textrm{max}\left[ \frac{(2+2 h +h^2) \textrm{e}^{\frac{2}{h}}}{h^2} \right], \label{C16}
\end{equation}
the maximum of this expression being taken in the region between the axis and the boundary. Moreover, from (\ref{finC9}), we see that $h'$ vanishes for $h=h_0=0$, and for $h_1$ such that $c_g = - \textrm{e}^{\frac{2}{h_1}}$. Therefore, $h$ exhibits finite values between the axis and the boundary and it is always possible to find negative values for $c_g$ verifying (\ref{C16}). Then, the strong energy condition is satisfied since $h \geq 0$.

The constraints issued from the axisymmetry and strong energy conditions, $h_0=0$ and $h>0$, are compatible provided that $h'>0$ in the vicinity of the axis. This implies that the sign in front of the square root issued from the integration of the $g(r)$ function and transferred to $h'$ in the course of the calculations is a plus sign, as implemented in the above equations describing the solutions and their main properties.

\subsubsection{Components of the four-velocity} \label{Cfv}

The $\Omega$ parameter diverges for $h_0=0$. However, this is not a drawback since this global rotation parameter is ill-defined on the very axis, considered as a line with no thickness, and therefore with no properly defined global rotation.

\subsubsection{Singularities}

The expressions for the metric functions $\textrm{e}^{\mu}$ and $f$, diverge for $h_0=0$ on the axis. 

Now, the energy density, and the pressure, vanish at the limit where $h \rightarrow 0$. Hence, the singularity on the axis is a mere coordinate singularity.

\subsubsection{Signature of the metric} \label{signax}

We see from (\ref{finC1}) that the metric function $\textrm{e}^\mu$ is positive for any value of $h$. Now, it has been shown in \cite{D96} that the metric signature is Lorentzian provided that $f$ and $l+k^2/f$ have the same sign as $\textrm{e}^\mu$. Moreover, we can write
\begin{equation}
l + \frac{k^2}{f} = \frac{D^2}{f}, \label{sign0}
\end{equation}
which implies that the sign of this expression is the same as that of $f$. Therefore, the constraint reduces to $f$ and $\textrm{e}^\mu$ exhibiting the same sign.

Owing to $\textrm{e}^{\mu} > 0$, we need $f > 0$, which is equivalent to
\begin{equation}
c_N \textrm{e}^{\frac{1}{h}} > \sqrt{2} c  \frac{(1+h)}{h \textrm{e}^{\frac{1}{h}}\left(\textrm{e}^{\frac{2}{h}} + c_g \right)} \left(c_k - \frac{1}{2^{\frac{3}{4}}} \sqrt{\frac{c_N}{c}} \textrm{e}^{\frac{2}{h}} \right)^2. \label{sign3}
\end{equation}
Since $c$, $c_N$, $h$ and $1+h$ are all positive, $c$ and $c_N$ from the occurrence of $\sqrt{c}$ and $\sqrt{c_N}$ in (\ref{finC8}) expressing the real quantity $\Omega$, and $h$, and thus $1+h$, from the strong energy condition, (\ref{sign3}) can be written as
\begin{equation}
\frac{c_N}{\sqrt{2} c} \frac{h}{(1+h)} \frac{\textrm{e}^{\frac{2}{h}}}{\left(c_k - \frac{1}{2^{\frac{3}{4}}} \sqrt{\frac{c_N}{c}} \textrm{e}^{\frac{2}{h}} \right)^2} > \left(\textrm{e}^{\frac{2}{h}} + c_g \right)^{-1} . \label{sign4}
\end{equation}
On the axis, where $h=0$, each side of (\ref{sign4}) tends to zero and this constraint is satisfied. Off the axis, the left hand side of (\ref{sign4}) is strictly positive and finite. It is therefore always possible, for a given set of ${c, c_N, c_k}$ parameters and a given value of $h$ on the boundary, to choose $c_g$ negative, with $-c_g \leq \textrm{max} \left( \textrm{e}^{\frac{2}{h}}\right)$, so that (\ref{sign4}) is verified. Class A possesses therefore a large subclass of solutions exhibiting a proper Lorentzian signature.

\section{Specialization to a fluid with purely azimuthal pressure} \label{azimuthal}

The equation of state applying to a fluid with azimuthally directed pressure is
\begin{equation}
P_z = P_r = 0. \label{eosaz}
\end{equation}

In this section, we consider two types of fluids, one where $h\equiv P_{\phi}/\rho$ depends on the radial coordinate $r$, one with $h= const$. A general method allowing to derive solutions of the first kind is displayed and a particular example is given. Then, the class $h=const$. is solved.

For the equation of state (\ref{eosaz}), the stress-energy tensor (\ref{setens}) becomes
\begin{equation}
T_{\alpha \beta} = \rho V_\alpha V_\beta + P_\phi K_\alpha K_\beta. \label{setensaz}
\end{equation}

Its conservation, implemented by the Bianchi identity (\ref{Bianchi2}), reads
\begin{equation}
(1+h) \frac{l'}{2l}- \frac{D'}{D} = 0. \label{Bianchi4az}
\end{equation}

\subsection{Field equations} \label{fea}

By inserting (\ref{partint8}) and (\ref{eosaz}) into (10)-(14) of \cite{CS20}, the five field equations become
\begin{equation}
G_{00} = \frac{\textrm{e}^{-\mu}}{2} \left[-f\mu'' - 2f\frac{D''}{D} + f'' - f'\frac{D'}{D} + \frac{3f(f'l' + k'^2)}{2D^2} \right] = \frac{\kappa}{l} \left(D^2 \rho + k^2 P_{\phi} \right), \label{G00az}
\end{equation}
\begin{equation}
G_{03} =  \frac{\textrm{e}^{-\mu}}{2} \left[k\mu'' + 2 k \frac{D''}{D} -k'' + k'\frac{D'}{D} - \frac{3k(f'l' + k'^2)}{2D^2}\right] = \kappa P_{\phi} k, \label{G03az}
\end{equation}
\begin{equation} 
G_{11} = \frac{\mu' D'}{2D} + \frac{f'l' + k'^2}{4D^2} = 0, \label{G11az}
\end{equation}
\begin{equation}
G_{22} = \frac{D''}{D} -\frac{\mu' D'}{2D} - \frac{f'l' + k'^2}{4D^2} = 0, \label{G22az}
\end{equation}
\begin{equation}
G_{33} =  \frac{\textrm{e}^{-\mu}}{2} \left[l\mu'' + 2l\frac{D''}{D} - l'' + l'\frac{D'}{D} - \frac{3l(f'l' + k'^2)}{2D^2}\right] =  \kappa P_{\phi} l. \label{G33az}
\end{equation}

\subsection{Solving the field equations for $h=h(r)$. General method} \label{solvingb}

The addition of both field equations (\ref{G11az}) and (\ref{G22az}) yields
\begin{equation}
D'' = 0, \label{sol1a}
\end{equation}
which can be integrated as
\begin{equation}
D = r+c_2, \label{sol3a}
\end{equation}
after rescaling of the radial coordinate $r$.

The problem being to solve seven equations, namely, (\ref{timelike}), (\ref{partint8}), (\ref{G00az})--(\ref{G33az}), for eight unknown functions of $r$, which are $f$, $k$, $\mu$, $l$, $v$, $\Omega$, $\rho$ and $P_{\phi}$, one more equation is needed to fully determine the system. A $l(h)$ function, designed such as to allow an easy integration of the Bianchi identity, is chosen as
\begin{equation}
l' = \frac{\textrm{d}l}{\textrm{d}h} h'. \label{solA}
\end{equation}
We insert (\ref{solA}) divided by $l(h)$ into the Bianchi identity (\ref{Bianchi4az}) and obtain
\begin{equation}
\frac{D'}{D} = \frac{(1+h)}{2}\frac{\frac{\textrm{d}l}{\textrm{d}h} h'}{l}, \label{solB}
\end{equation}
which is designed such as to be also integrable with respect to $r$ under the form 
\begin{equation}
D=D(h). \label{solC}
\end{equation}
Now, by equalizing both expressions for $D$, (\ref{sol3a}) and (\ref{solC}), we obtain an equation for $h$ as a function of $r$ under the form
\begin{equation}
r+c_2 = D(h). \label{solD}
\end{equation}
By differentiating (\ref{solD}) with respect to $r$, we obtain
\begin{equation}
h'(h) = \frac{1}{\frac{\textrm{d}D(h)}{\textrm{d}h}}. \label{solE}
\end{equation}

Then, by substituting (\ref{partint14}) into (\ref{G11az}), we obtain
\begin{equation}
\frac{\mu' D'}{D} = -\frac{l'D'}{lD} + \frac{l'^2}{2l^2} - \frac{2c^2}{l^2}, \label{partint15}
\end{equation}
where we substitute (\ref{solA}), (\ref{solB}), and $l(h)$ which gives
\begin{equation}
\mu' = {\frac{\textrm{d}\mu(h)}{\textrm{d}h}}h'. \label{solF}
\end{equation}
If the choice of $l(h)$ has been judiciously made, $\textrm{d}\mu(h) /\textrm{d}h$ happens to be an integrable function of $h$ and, therefore, $\mu(h)$ follows. Knowing $D(h)$, $l(h)$, and $h'(h)$, the metric function $k$ proceeds from (\ref{partint12}). Then $f$ is obtained from the definition (\ref{D2}) of $D^2$.

The field equation (\ref{G33az}), with (\ref{G11az}), (\ref{G22az}), the metric functions, $D$ and derivatives inserted, yields the expression for the pressure $P_{\phi}$, from which $\rho= P_{\phi}/h$ proceeds. The differential angular velocity $\Omega$ and the velocity $v$ of the fluid follow from (\ref{partint8}) and (\ref{partint9}).

To exemplify this method, a particular class of solutions is considered in Sec. \ref{solvinga}.

\subsection{Solving the field equations for $h=h(r)$. Class 1 example} \label{solvinga}

To use the left degree of freedom, a particular form of the $l$ function is chosen as 
\begin{equation}
l = c_l^2 \frac{h^2}{(1-h)^2}. \label{sol5a}
\end{equation}
This particular expression for $l$ has been chosen to exemplify our method since it allows to obtain a fully analytic and physically well-behaved solution with a rather simple form and, thus, a straightforward interpretation.

By implementing the method described in Sec.\ref{solvingb}, we obtain the following solution:
\begin{equation}
l = c_l^2 \frac{h^2}{(1-h)^2}, \label{fin1}
\end{equation}
\begin{equation}
\textrm{e}^{\mu} = \frac{(1-h)^{\frac{4c^4}{c_l^4} +1}}{h^{\frac{4c^4}{c_l^4}}(1+h)} \exp\left[\frac{2c^4}{c_l^4}\left(\frac{2}{h} + \frac{1}{2h^2}\right)\right], \label{fin2}
\end{equation}
\begin{equation}
k = c_l^2 \frac{h^2}{(1-h)^2} \left[c_k - \frac{2c^2}{c_l^4} \left(\frac{2}{h} + \frac{1}{2h^2}+ \ln \frac{(1-h)^2}{h^2} \right) \right], \label{fin3}
\end{equation}
\begin{equation}
f = \frac{1}{c_l^2 (1-h)^2} - c_l^2 \frac{h^2}{(1-h)^2}\left[c_k  - \frac{2c^2}{c_l^4} \left(\frac{2}{h} + \frac{1}{2h^2} + \ln \frac{(1-h)^2}{h^2} \right) \right]^2,  \label{fin4}
\end{equation}
\begin{equation}
D = \frac{h}{(1-h)^2} = r, \label{fin5}
\end{equation}
\begin{equation}
h = 1 +\frac{1}{2r} - \sqrt{\frac{1}{r} + \frac{1}{4r^2}}, \label{fin6}
\end{equation}
\begin{equation}
h' = \frac{(1-h)^3}{(1+h)}= -\frac{1}{2r^2}\left(1 - \frac{1 + 2r}{\sqrt{1+4r}} \right), \label{fin7}
\end{equation}
\begin{equation}
\rho =\frac{2 h^{\frac{4c^4}{c_l^4} -1}}{\kappa (1-h)^{\frac{4c^4}{c_l^4} -3}} \left[\frac{(1-h)}{(1+h)^3} - \frac{2c^4}{c_l^4 h^3}\right] \exp\left[-\frac{2c^4}{c_l^4}\left(\frac{2}{h} + \frac{1}{2h^2}\right)\right], \label{fin8}
\end{equation}
\begin{equation}
P_{\phi} =\frac{2 h^{\frac{4c^4}{c_l^4}}}{\kappa (1-h)^{\frac{4c^4}{c_l^4} -3}} \left[\frac{(1-h)}{(1+h)^3} - \frac{2c^4}{c_l^4 h^3}\right] \exp\left[-\frac{2c^4}{c_l^4}\left(\frac{2}{h} + \frac{1}{2h^2}\right)\right], \label{fin8a}
\end{equation}
\begin{equation}
v = c_l (1-h), \label{fin9}
\end{equation}
\begin{equation}
\Omega = -c_l (1-h)  \left[c_k - \frac{2c^2}{c_l^4} \left(\frac{2}{h} + \frac{1}{2h^2} + \ln \frac{(1-h)^2}{h^2} \right) \right], \label{pfin10}
\end{equation}
\begin{equation}
\dot{V^{\alpha}}\dot{V_{\alpha}} = \frac{h^{\frac{4c^4}{c_l^4}}}{(1-h)^{\frac{4c^4}{c_l^4} -3}(1+h)}\exp\left[-\frac{2c^4}{c_l^4}\left(\frac{2}{h} + \frac{1}{2h^2}\right)\right], \label{fin11}
\end{equation}
\begin{equation}
\omega =0, \label{fin12}
\end{equation}
\begin{equation}
\sigma^2= \frac{c^4}{c_l^4} \frac{h^{4\left(\frac{c^4}{c_l^4} -1\right)}(1+h)}{(1-h)^{\frac{4c^4}{c_l^4} -3}} \exp\left[-\frac{2c^4}{c_l^4}\left(\frac{2}{h} + \frac{1}{2h^2}\right)\right]. \label{fin13}
\end{equation}

\subsubsection{Behaviour of the $h(r)$ function and energy conditions} \label{signconsaz}

To satisfy the axisymmetry condition, $l\stackrel{0}{=}0$ \cite{C23a,S09}, where $\stackrel{0}{=}$ denotes quantities evaluated in the limit at the axis of symmetry, the value of $h$ at $r=0$ is
\begin{equation}
h_0 = 0. \label{arc2}
\end{equation}

In the course of the calculations made to obtain the exact expressions for the solutions displayed above, $h(r)$ was initially obtained as
\begin{equation}
h = 1 +\frac{1}{2r} +\epsilon{\frac{1}{r} + \frac{1}{4r^2}}, \label{fin6}
\end{equation}
with $\epsilon = \pm1$. Therefore, we have $h(r) >0$ whatever the sign of $\epsilon$. 

The first derivative of $h(r)$ is unity on the axis where $h_0=0$ and vanishes for $h=+1$ which is the upper limit of the definition interval of $h$. It is therefore positive in the full interval. Hence, the function $h$ keeps growing from $h_0=0$ towards its maximum, implying that $h$ is everywhere positive and that its interval of definition is
\begin{equation}
0<h<+1. \label{sol22b}
\end{equation}
With this constraint accounted for, the weak energy condition, $\rho>0$, imposes
\begin{equation}
2 \frac{c^2}{c_l^4}<\frac{h^3 (1-h)}{(1+h)^3}. \label{sol22bb}
\end{equation}
A straightforward mathematical analysis of the expression at the right hand side of (\ref{sol22bb}) shows that its first derivative with respect to $h$ vanishes for two numerical values located outside the interval of definition (\ref{sol22b}). This derivative keeps thus the same positive sign at all points of the interior spacetimes. The corresponding expression is therefore monotonically increasing from the axis, where it vanishes, towards the boundary. Thus, for any couple of parameters $\{c,c_l\}$, the weak energy condition is verified at all points where $h$ satisfies $h_1<h<h_{\Sigma}<1$, with $h_1$ defined by
\begin{equation}
2 \frac{c^2}{c_l^4}=\frac{h_1^3 (1-h_1)}{(1+h_1)^3}. \label{sol22c}
\end{equation}
The larger $c_l$ with respect to $c$, the closer from the axis this condition is satisfied.  The strong energy condition follows from $h>0$.

Then, considering the first derivative of the function $h(r)$ written as
\begin{equation}
h' = -\frac{1}{2r^2}\left(1 + \epsilon\frac{1 + 2r}{\sqrt{1+4r}} \right). \label{sol22aa}
\end{equation}
Any well-behaved solution implies $\epsilon = -1$, which is the choice made above in (\ref{fin1})-(\ref{fin13}). Recall that (\ref{sol22bb}) and (\ref{sol22aa}) apply only outside the axis, i.e., for $r \neq 0$, since they are obtained after dividing by $r$ during the calculations.

\subsubsection{Metric signature}

To preserve the Lorentzian signature of the metric, we have already seen in Sec.\ref{signax} that the metric functions $f$ and $\textrm{e}^\mu$ must have the same sign. For $h$ fulfilling constraint (\ref{sol22b}), $\textrm{e}^\mu$ exhibits a positive sign. Hence $f$ should be positive definite.

Now, owing to (\ref{fin4}), a positive metric function $f$ is obtained if and only if
\begin{equation}
1 > c_k c_l^2h + \frac{2c}{c_l^2} \left[2 + \frac{1}{2h} +h \ln \frac{(1-h)^2}{h^2}\right]. \label{sol25b}
\end{equation}
At both extrema of the definition interval of $h$, $\{0,+1\}$, the term inside the brackets diverges. Thus, this inequality can be fulfilled only if $c<0$. Hence, $c$ will be denoted $-c^2$ in the following. Moreover, either $c_k$ is negative and the inequality is verified without any other requirement, or $c_k$ is positive and (\ref{sol25b}) is a constraint on the three parameters $c^2$, $c_l^2$ and $c_k$, depending on the value of $h_{\Sigma}$ on the boundary, since the maximum $h=1$ is not reached inside the cylinder of fluid, as shown in Sec.\ref{signconsaz}.

\subsubsection{Singularities}

The solutions displayed here seem, at first sight, to involve two possible singular loci.

One might occur for $h=+1$ where the metric functions diverge. The radial coordinate $r$, given by (\ref{fin5}), diverges also for this value of $h$. However, since $r$ is bounded by the radius of the cylinder $r_{\Sigma}$, $h=+1$, is never reached inside the fluid.

Another singularity seems to occur for $h=r=0$, i.e., on the axis. For such a value of $h$, the metric functions $k$, $\textrm{e}^{\mu}$ and $l$ vanish and the density diverges. However, this is a mere mathematical decoy, since it emerges from repeated divisions by $h$  during the calculations. For instance, in (\ref{fin6}), $h$ diverges for $r=0$, instead of vanishing.

\subsection{Solving the field equations for $h=const.$ Class 2} \label{solvingc}

We consider now the case where the ratio $h$ is a constant, which is known to have a number of physical applications in astrophysics. The beginning of the calculations, which is independent of $h$, is the same as in Sec.\ref{solvingb} and yields similarly (\ref{sol1a})-(\ref{sol3a}).

Then, we insert (\ref{sol3a}) into the Bianchi identity (\ref{Bianchi4az}) and obtain
\begin{equation}
\frac{l'}{l} = \frac{2}{(1+h)(r+c_2)}, \label{sol1b}
\end{equation}
where $c_2$ is a constant of integration, and which can be integrated to give
\begin{equation}
l = c_3 (r+ c_2)^{\frac{2}{1+h}}, \label{sol2b}
\end{equation}
where $c_3$ is an integration constant.

The axisymmetry condition $l\stackrel{0}{=}0$ is satisfied provided that $c_2 = 0$, which implies
\begin{equation}
l = c_3 r^{\frac{2}{1+h}}, \label{sol3b}
\end{equation}
and
\begin{equation}
D = r. \label{sol4b}
\end{equation}

By implementing (\ref{sol3b}) and (\ref{sol4b}) into (\ref{partint12}), we obtain
\begin{equation}
k = c_3 r^{\frac{2}{1+h}} \left[c_k + \frac{c(1+h)}{c_3^2(1-h)} r^{\frac{2(h-1)}{h+1}} \right], \label{sol5b}
\end{equation}
where $c_k$ is another integration constant. The metric function $f$ follows from (\ref{D2}), as
\begin{equation}
f = \frac{r^{\frac{2h}{1+h}}}{c_3} - c_3 r^{\frac{2}{1+h}} \left[c_k + \frac{c(1+h)}{c_3^2(1-h)} r^{\frac{2(h-1)}{h+1}} \right]^2. \label{sol6b}
\end{equation}

By using (\ref{G11az}), we obtain
\begin{equation}
\mu' = - \frac{2h}{(1+h)^2 r} - \frac{2 c^2}{c_3^2} r^{\frac{h-3}{h+1}}. \label{sol7b}
\end{equation}

There are two classes of solutions to (\ref{sol7b}), depending on the value taken by $h$.

\subsubsection{Subclass i: $h \neq 1$} \label{classA}

When $h \neq 1$, and thus $(h-3)/(h+1) \neq -1$, $\mu'$ can be integrated such as to give
\begin{equation}
\textrm{e}^{\mu} =  \frac{1}{r^{\frac{2h}{(1+h)^2}}} \exp\left[ \frac{ c^2(1+h)}{c_4^2(1-h)} r^{\frac{2(h-1)}{h+1}} \right], \label{sol8b}
\end{equation}
where $c_4$ is a new constant of integration.

Then, the calculations are performed as previously and we obtain
\begin{equation}
l = c_4 r^{\frac{2}{1+h}}, \label{fin1b}
\end{equation}
\begin{equation}
\textrm{e}^{\mu} = \frac{1}{r^{\frac{2h}{(1+h)^2}}} \exp\left[ \frac{ c^2(1+h)}{c_4^2(1-h)r^{\frac{2(1-h)}{1+h}}}  \right], \label{fin2b}
\end{equation}
\begin{equation}
k = \frac{c(1+h)}{c_4(1-h)} r^{\frac{2h}{1+h}}, \label{fin3b}
\end{equation}
\begin{equation}
f = \frac{r^{\frac{2h}{1+h}}}{c_4} - \frac{c^2(1+h)^2}{c_4^3(1-h)^2} r^{\frac{2(2h-1)}{1+h}}, \label{fin4b}
\end{equation}
\begin{equation}
D = r, \label{fin5b}
\end{equation}
\begin{equation}
\rho = - \frac{4 c^2}{\kappa c_4^2 (1+h) r^{\frac{2(2+h)}{(1+h)^2}}} \exp\left[ - \frac{ c^2(1+h)}{c_4^2(1-h) r^{\frac{2(1-h)}{1+h}}} \right], \label{fin6b}
\end{equation}
\begin{equation}
P_{\phi} = - \frac{4 c^2 h}{\kappa c_4^2 (1+h) r^{\frac{2(2+h)}{(1+h)^2}}} \exp\left[ - \frac{ c^2(1+h)}{c_4^2(1-h) r^{\frac{2(1-h)}{1+h}}} \right], \label{fin7b}
\end{equation}
\begin{equation}
v = \frac{\sqrt{c_4}}{r^{\frac{h}{1+h}}}, \label{fin8b}
\end{equation}
\begin{equation}
\Omega = - \frac{c(1+h)}{c_4^{\frac{3}{2}}(1-h) r^{\frac{2-h}{1+h}}}, \label{fin9b}
\end{equation}
\begin{equation}
\dot{V^{\alpha}}\dot{V_{\alpha}} = \frac{h^2}{(1+h)^2} r^{\frac{2h}{(1+h)^2} - 2} \exp\left[-\frac{c^2(1+h)}{c_4^2(1-h) r^{\frac{2(1-h)}{1+h}}}\right], \label{fin10b}
\end{equation}
\begin{equation}
\omega = 0, \label{fin11b}
\end{equation}
\begin{equation}
\sigma^2= \frac{c^2}{c_4^2 r^{\frac{2(2+h)}{(1+h)^2}}} \exp\left[-\frac{c^2(1+h)}{c_4^2(1-h) r^{\frac{2(1-h)}{1+h}}}\right]. \label{fin12b}
\end{equation}

\hfill

{\bf $\alpha$)} {\it Energy density}

\hfill

The energy density vanishes at the axis where $r=0$, and such does the pressure. Therefore, the axis of rotation is nonsingular.

Moreover, we have $\rho >0$ if $h< -1$, which implies $P_{\phi}<0$. Since a negative pressure is encountered in various physical configurations \cite{C23c}, this is acceptable.

\hfill

{\bf $\beta$)} {\it Singularities}

\hfill

No metric function diverges for $h \neq 1$, hence the solutions are singularity-free.

\hfill

{\bf $\gamma$)} {\it Metric signature}

\hfill

The metric function $\textrm{e}^{\mu}$, as given by (\ref{fin2b}), is obviously positive. The function $f$, initially written as
\begin{equation}
f = \frac{r^{\frac{2h}{1+h}}}{c_4} - c_4 r^{\frac{2}{1+h}} \left[c_k + \frac{c(1+h)}{c_4^2(1-h)} r^{\frac{2(h-1)}{h+1}} \right]^2, \label{sol6b}
\end{equation}
must also be positive.

A straightforward analysis of this expression shows that $f>0$ implies
\begin{eqnarray}
\frac{1 - 2c c_k \frac{1+h}{1-h} - \sqrt{\left[2 c c_k \frac{1+h}{1-h} - 1\right]^2 - 4 c^2 c_k^2 c_4 \frac{(1+h)^2}{(1-h)^2}}}{\frac{2 c^2 (1+h)^2}{c_4 (1-h)^2}} < r^{\frac{2(h-1)}{h+1}} \nonumber \\
< \frac{1 - 2c c_k \frac{1+h}{1-h} + \sqrt{\left[2 c c_k \frac{1+h}{1-h} - 1\right]^2 - 4 c^2 c_k^2 c_4 \frac{(1+h)^2}{(1-h)^2}}}{\frac{2 c^2 (1+h)^2}{c_4 (1-h)^2}}. \label{sol19b}
\end{eqnarray}

We must consider two cases:

\hfill

Case (a): $\frac{h-1}{h+1} > 0$, which corresponds either to $h>1$ or to $h< -1$. In this case, $r^{\frac{2(h-1)}{h+1}}$ is an increasing function of $r$. For $r=0$, we can write
\begin{equation}
\left[1 - 2c c_k \frac{1+h}{1-h}\right]^2 = \left[1 - 2c c_k \frac{1+h}{1-h}\right]^2 - 4 c^2 c_k^2 c_4 \frac{(1+h)^2}{(1-h)^2}. \label{sol200b}
\end{equation}
This equality imposes that the last term on the right-hand side vanishes. This can be performed through the vanishing of one of the three parameters occurring there. However, since only the vanishing of $c_k$ does not cause any damage to the solutions, we choose $c_k=0$ to deal properly with the metric signature, in this case.

\hfill

Case (b): $\frac{h-1}{h+1} < 0$, which corresponds to $-1<h<+1$. Thus, $r^{\frac{2(h-1)}{h+1}}$ is a decreasing function of $r$, which gives, for $r \rightarrow 0$, $r^{\frac{2(h-1)}{h+1}} \rightarrow \infty$. This is a pathological feature which rules out case (b).

\subsubsection{Subclass ii: $h=1$}

By substituting $h=1$ into (\ref{sol7b}), we obtain, after integration,
\begin{equation}
\textrm{e}^{\mu} = \frac{1}{r^{\frac{1}{2}+\frac{2 c^2}{c_3^2}}}. \label{sol16b}
\end{equation}

Now, $l$ given by (\ref{sol3b}), becomes, with $h=1$,
\begin{equation}
l = c_3 r, \label{sol16ab}
\end{equation}
which, inserted into (\ref{partint12}), together with (\ref{sol4b}), gives
\begin{equation}
k = c_3 r \left(c_k - \frac{2c}{c_3^2} \ln r \right). \label{sol16bb}
\end{equation}
Then, the metric function $f$ follows from (\ref{D2}), and reads
\begin{equation}
f = \frac{r}{c_3} - c_3 r \left(c_k - \frac{2c}{c_3^2} \ln r \right)^2. \label{sol6cb}
\end{equation}

The pressure can be calculated by using (\ref{G33az}), which yields, with $h=1$,
\begin{equation}
\rho = P_{\phi} = - \frac{2 c^2}{\kappa c_3^2} r^{\frac{2 c^2}{c_3 ^2} - \frac{3}{2}}. \label{sol18b}
\end{equation}
The energy density is negative which compromises the behaviour of these solutions.

\section{Specialization to a fluid with purely radial pressure} \label{radial}

The equation of state for an anisotropic fluid with radially directed pressure is
\begin{equation}
P_z = P_{\phi} = 0. \label{eosr}
\end{equation}

One degree of freedom is therefore still left. As it has been the case when dealing with rigidly rotating such fluids, a combination of the field equations can be factorized. Here, four factors appear, each defining a given class of solutions. The remaining degree of freedom is thus used to make a choice among these classes.

For a radially directed pressure, the stress-energy tensor (\ref{setens}) reads
\begin{equation}
T_{\alpha \beta} = (\rho + P_r) V_\alpha V_\beta + P_r g_{\alpha \beta} - P_r K_\alpha K_\beta - P_r S_\alpha S_\beta. \label{setens2r}
\end{equation}

The Bianchi identity (\ref{Bianchi2}), becomes
\begin{equation}
\frac{P'_r}{P_r} - \frac{l'}{2hl}+\frac{(1+h)}{h} \frac{D'}{D} + \frac{\mu'}{2}= 0. \label{Bianchi4r}
\end{equation}

\subsection{Field equations} \label{fea}

By inserting the components of the stress-energy tensor and the equations of state into (10)-(14) of \cite{CS20}, the five field equations can be written as
\begin{equation}
G_{00} = \frac{\textrm{e}^{-\mu}}{2} \left[-f\mu'' - 2f\frac{D''}{D} + f'' - f'\frac{D'}{D} + \frac{3f(f'l' + k'^2)}{2D^2} \right] = \frac{\kappa \rho D^2}{l}, \label{G00r}
\end{equation}
\begin{equation}
G_{03} =  \frac{\textrm{e}^{-\mu}}{2} \left[k\mu'' + 2 k \frac{D''}{D} -k'' + k'\frac{D'}{D} - \frac{3k(f'l' + k'^2)}{2D^2}\right] = 0, \label{G03r}
\end{equation}
\begin{equation} 
G_{11} = \frac{\mu' D'}{2D} + \frac{f'l' + k'^2}{4D^2} = \kappa P_r \textrm{e}^{\mu}, \label{G11r}
\end{equation}
\begin{equation}
G_{22} = \frac{D''}{D} -\frac{\mu' D'}{2D} - \frac{f'l' + k'^2}{4D^2} = 0, \label{G22r}
\end{equation}
\begin{equation}
G_{33} =  \frac{\textrm{e}^{-\mu}}{2} \left[l\mu'' + 2l\frac{D''}{D} - l'' + l'\frac{D'}{D} - \frac{3l(f'l' + k'^2)}{2D^2}\right] =  0. \label{G33r}
\end{equation}

\subsection{Identification of the four classes of solutions} \label{identif}

As it has been the case when dealing with rigidly rotating fluids with radially directed pressure \cite{C23d}, a combination of the field equations can be factorized. Here, four factors appear again, each defining a given class of solutions. They are obtained as follows.

The addition of both field equations (\ref{G11r}) and (\ref{G22r}) yields
\begin{equation}
\frac{D''}{D} = \kappa P_r \textrm{e}^{\mu}, \label{fac1}
\end{equation}
which we insert into (\ref{G22r}) together with (\ref{partint14}) so as to obtain
\begin{equation}
\left(\mu' + \frac{l'}{l}\right)\frac{D'}{D} - \frac{l'^2}{2l^2} + \frac{2c^2}{l^2} = 2 \kappa P_r \textrm{e}^{\mu}, \label{fac2}
\end{equation}
which we differentiate with respect to $r$ which gives
\begin{eqnarray}
\left(\mu'' + \frac{l''}{l} - \frac{l'^2}{l^2} \right)\frac{D'}{D} &+& \left(\mu' + \frac{l'}{l}\right)\left(\frac{D''}{D} - \frac{D'^2}{D^2} \right) - \frac{l' l''}{l^2} + \frac{l'^3}{l^3} - \frac{4c^2 l'}{l^3} \nonumber \\
&=&  2 \kappa P'_r \textrm{e}^{\mu} + 2 \kappa P_r \textrm{e}^{\mu} \mu', \label{fac3}
\end{eqnarray}
where we insert $P'_r$ as given by (\ref{Bianchi4r}), and $D''/D$ as given by (\ref{fac1}) so as to obtain
\begin{eqnarray}
&& \left(\mu'' + \frac{l''}{l} - \frac{l'^2}{l^2} \right)\frac{D'}{D} - \left(\mu' + \frac{l'}{l}\right)\frac{D'^2}{D^2} - \frac{l' l''}{l^2} + \frac{l'^3}{l^3} - \frac{4c^2 l'}{l^3} \nonumber \\
&=&  \kappa P_r \textrm{e}^{\mu} \left[\frac{(1-h)}{h} \frac{l'}{l} - \frac{2(1+h)}{h} \frac{D'}{D} \right]. \label{fac4}
\end{eqnarray}

Now, we write (\ref{fac2}) as
\begin{equation}
\frac{D'}{D} = \frac{2}{\mu' + \frac{l'}{l}}\left( \frac{l'^2}{4l^2} - \frac{c^2}{l^2} + \kappa P_r \textrm{e}^{\mu}\right), \label{fac5}
\end{equation}
which we insert into (\ref{fac4}) and obtain
\begin{eqnarray}
&-& \frac{4 \kappa^2 P_r^2 \textrm{e}^{2\mu}}{h \left(\mu' + \frac{l'}{l} \right)} + \frac{\kappa P_r \textrm{e}^{\mu}}{\mu' + \frac{l'}{l}} \left[-\frac{4(1-h)}{h} \left( \frac{l'^2}{4l^2} - \frac{c^2}{l^2}\right) - 2 \left(\mu'' + \frac{l''}{l} - \frac{l'^2}{l^2} \right) \right. \nonumber \\
&&+ \left. \frac{(1-h)}{h} \frac{l'}{l} \left(\mu' + \frac{l'}{l} \right) \right] + \frac{2}{\mu' + \frac{l'}{l}} \left( \frac{l'^2}{4l^2} - \frac{c^2}{l^2}\right) \left[2 \left( \frac{l'^2}{4l^2} - \frac{c^2}{l^2}\right) \right. \nonumber \\
&& \left. - \mu'' - \frac{l''}{l} + \frac{l'^2}{l^2} \right] 
+ \frac{l' l''}{l^2} - \frac{l'^3}{l^3} + \frac{4c^2 l'}{l^3} =0. \label{fac6}
\end{eqnarray}

Then, we write (\ref{G22r}) as
\begin{equation}
\frac{f'l' + k'^2}{2D^2} = \frac{2D''}{D} - \frac{\mu' D'}{D}, \label{fac7}
\end{equation}
which we insert into (\ref{G33r}) so as to obtain
\begin{equation}
\frac{D'}{D} = \frac{- \mu'' + \frac{l''}{l} + 4 \kappa P_r \textrm{e}^{\mu}}{3 \mu' + \frac{l'}{l}}. \label{fac8}
\end{equation}
Then, we equalize (\ref{fac5}) and (\ref{fac8}), which gives
\begin{equation}
\kappa P_r \textrm{e}^{\mu} = - \left[\frac{2 \left(3 \mu' + \frac{l'}{l}\right) \left(\frac{l'^2}{4 l^2} - \frac{c^2}{l^2}\right) + \left(\mu' + \frac{l'}{l}\right) \left(\mu'' - \frac{l''}{l}\right)}{2 \left(\mu' - \frac{l'}{l}\right)} \right], \label{fac9}
\end{equation}
which we substitute into (\ref{fac6}) and obtain, after some arrangements,
\begin{eqnarray}
&& \left(\mu' + \frac{l'}{l}\right)\left[(3+h)\mu' + (1-h)\frac{l'}{l} \right] \left(2\mu'' - \frac{2l''}{l} + \frac{\mu' l'}{l} + \frac{l'^2}{l^2} - \frac{8 c^2}{l^2} \right) \nonumber \\
&\times& \left\{ \left[(3 + h)\mu' +(1-h)\frac{l'}{l} \right]  \left( \frac{4c^2}{l^2} - \frac{l'^2}{l^2} \right) - 2 \left[(1 + h)\mu' +(1-h)\frac{l'}{l} \right] \right. \nonumber \\
&\times& \left.\left(\mu'' - \frac{l''}{l}\right)\right\} = 0. \label{fac10}
\end{eqnarray}
It corresponds to four classes of solutions. We denote them as class I, II, III, and IV, defined by the vanishing of the first, second, third, and forth factor, respectively. To proceed, we will need the following equations, also obtained from the field equations:
\begin{equation}
\frac{D'}{D} = - \frac{\mu'' - \frac{l''}{l} + \frac{l'^2}{l^2} -\frac{4c^2}{l^2}}{\mu' - \frac{l'}{l}}, \label{IIIb}
\end{equation}
\begin{equation}
\mu'' -  \frac{4 D''}{D} - \frac{l''}{l}  + \left(3\mu' + \frac{l'}{l} \right) \frac{D'}{D} = 0, \label{IIIg}
\end{equation}
\begin{equation}
\frac{D''}{D} = \left(\mu' + \frac{l'}{l} \right)\frac{D'}{2D} - \frac{l'^2}{4l^2} + \frac{c^2}{l^2}. \label{plus1}
\end{equation}

\subsection{Class I} \label{cI}

As specified above, the defining equation of this class of solutions is
\begin{equation}
\mu' + \frac{l'}{l} = 0, \label{Ia}
\end{equation}
which can be integrated by
\begin{equation}
\textrm{e}^{\mu} =  \frac{1}{l}. \label{Ic}
\end{equation}

Then, by inserting (\ref{Ia}) into (\ref{fac2}), we obtain
\begin{equation}
2 \kappa P_r \textrm{e}^{\mu} = \frac{2c^2}{l^2} - \frac{l'^2}{2l^2}. \label{Id}
\end{equation}
Now, (\ref{Ia}) and (\ref{Id}) inserted into (\ref{G11r}) yield
\begin{equation}
\frac{f'l' + k'^2}{2D^2} = \frac{l' D'}{l D} - \frac{l'^2}{2l^2} +  \frac{2c^2}{l^2}. \label{Ie}
\end{equation}
By substituting (\ref{Id}) into (\ref{fac1}), we obtain
\begin{equation}
\frac{D''}{D} = \frac{c^2}{l^2} - \frac{l'^2}{4l^2}. \label{If}
\end{equation}
Then, we insert (\ref{Ia}) and its derivative, (\ref{Ie}) and (\ref{If}) into (\ref{G33r}), which yields
\begin{equation}
\frac{D'}{D} = - \frac{l''}{l'} +  \frac{l'}{l} - \frac{2c^2}{ll'}. \label{Ig}
\end{equation}
The metric function $\textrm{e}^{\mu}$ as given by (\ref{Ic}) is now inserted into (\ref{Id}) so as to obtain
\begin{equation}
P_r  =  \frac{1}{2 \kappa} \left(\frac{2c^2}{l} - \frac{l'^2}{2l} \right), \label{Ih}
\end{equation}
which, after differentiation with respect to $r$, can be inserted, together with (\ref{Ia}) and (\ref{Ig}), into the Bianchi identity (\ref{Bianchi4r}), in order to obtain
\begin{equation}
- \frac{l''}{l'} \left(\frac{4c^2 + l'^2}{4c^2-l'^2} + \frac{1}{h} \right) + \frac{l'^2 - 4c^2}{2hll'} - \frac{l'^2 + 4c^2}{2ll'} = 0. \label{Ij}
\end{equation}

To make the calculations easier, a function $p(r)$ is defined as
\begin{equation}
p = \frac{4c^2}{l'^2}.  \label{Ik}
\end{equation}
Now, (\ref{Ik}) and its derivative are inserted into (\ref{Ij}), which can thus be integrated by
\begin{equation}
l =c_p \frac{1-p}{p},  \label{In}
\end{equation}
where $c_p$ is an integration constant. By substituting (\ref{Ik}) into (\ref{In}), we obtain
\begin{equation}
l'^2 = 4c^2 \left(\frac{l}{c_p} +1 \right),  \label{Io}
\end{equation}
which can be inserted into (\ref{Ih}) so as to give
\begin{equation}
P_r  =  - \frac{c^2}{\kappa c_p}. \label{Ip}
\end{equation}

Now, (\ref{Io}) possesses two solutions, distinguished by $\epsilon = \pm 1$, and which read
\begin{equation}
l = \frac{c^2 c_5^2}{c_p} - c_p +\frac{2 \epsilon c^2 c_5}{c_p} r + \frac{c^2}{c_p}r^2,  \label{Iq}
\end{equation}
where $c_5$ is a constant of integration. 

The axisymmetry condition imposes the constraint:
\begin{equation}
c_p  =  c c_5. \label{Ir}
\end{equation}
The expression for $l$ simplifies, therefore, as
\begin{equation}
l = c r \left(2 \epsilon + \frac{r}{c_5}\right).  \label{Is}
\end{equation}

Then, we insert (\ref{Ia}) and (\ref{Ip}) into (\ref{Bianchi4r}) and integrate such as to obtain
\begin{equation}
D = c_6 \sqrt{l}, \label{Iu}
\end{equation}
where $c_6$ is another integration constant. By inserting (\ref{Is}) into (\ref{Iu}), we obtain
\begin{equation}
D = c_6 \sqrt{ c r \left(2 \epsilon + \frac{r}{c_5}\right)}, \label{Iv}
\end{equation}
and, analogously, into (\ref{Ic}),
\begin{equation}
\textrm{e}^{\mu} =  \frac{1}{ c r \left(2 \epsilon + \frac{r}{c_5}\right)}. \label{Iw}
\end{equation}

Now, we substitute (\ref{Is}) and (\ref{Iv}) into (\ref{partint12}) and obtain, after integration,
\begin{equation}
k = c r \left(2 \epsilon + \frac{r}{c_5}\right) \left[c_k + \frac{2 c_6\left(\epsilon + \frac{r}{c_5}\right)}{\sqrt{ c r \left(2 \epsilon + \frac{r}{c_5}\right)}}\right]. \label{Ix}
\end{equation}

Then $f(r)$ follows, as usual, through (\ref{D2}), which gives
\begin{equation}
f = c_6^2 - c r \left(2 \epsilon + \frac{r}{c_5}\right) \left[c_k + \frac{2 c_6\left(\epsilon + \frac{r}{c_5}\right)}{\sqrt{ c r \left(2 \epsilon + \frac{r}{c_5}\right)}}\right]^2. \label{Iy}
\end{equation}

Subsequently, the energy density $\rho$ is obtained from (\ref{G00r}), and reads
\begin{eqnarray}
\rho &=& \frac{3 c}{\kappa c_6^2 c_l r \left(2 \epsilon + \frac{r}{c_5}\right)} \left\{\left[c_5 +  2 \epsilon(1 - c c_5)r + \left(\frac{1}{c_5} - 5 c \right) r^2 \right. \right. \nonumber \\
&-& \left. \left. \frac{4\epsilon c}{c_5} r^3 - \frac{c}{c_5^2} r^4 \right] \left[c_k + \frac{2 c_6\left(\epsilon + \frac{r}{c_5}\right)}{\sqrt{ c r \left(2 \epsilon + \frac{r}{c_5}\right)}}\right]^2 - c_6^2 c_5 \right\}. \label{Iz}
\end{eqnarray}

By inserting (\ref{Is}) and (\ref{Iv}) into (\ref{partint9}), the velocity of the fluid follows as
\begin{equation}
v = \frac{1}{c_6}. \label{Iab}
\end{equation}
Then, the differential rotation parameter $\Omega$ is obtained from (\ref{partint8}) as
\begin{equation}
\Omega = - \frac{1}{c_6} \left[c_k + \frac{2 c_6\left(\epsilon + \frac{r}{c_5}\right)}{\sqrt{ c r \left(2 \epsilon + \frac{r}{c_5}\right)}}\right]. \label{Iac}
\end{equation}

We see that $\Omega$ diverges on the axis where $r=0$. However, as stressed in Sec.\ref{Cfv}, this feature is not a drawback since $\Omega$ is ill-defined on the axis.

\subsubsection{No angular deficit condition}

The so-called ``regularity'' condition, which rather avoids the occurrence of an angular deficit in the vicinity of the axis (see Appendix \ref{secA1}), reads
\begin{equation}
\frac{\textrm{e}^{-\mu}l'^2}{4l} \stackrel{0}{=} 1. \label{Iad}
\end{equation}
With (\ref{Is}) and (\ref{Iw}) inserted in (\ref{Iad}), this condition becomes
\begin{equation}
c = \epsilon. \label{Iae}
\end{equation}

Contrary to what occurs in the cases of axially and azimuthally directed pressure, class I where the pressure is radial does satisfy the condition for no angular deficit near the axis. We will henceforth apply constraint (\ref{Iae}) on the $c$ integration constant.

\subsubsection{Real number condition}

We impose that the mathematical expressions for the metric and the various attached physical quantities should be written with real numbers only. This implies
\begin{equation}
2 \epsilon + \frac{r}{c_5} > 0  \qquad \textrm{and} \qquad  2 + \frac{\epsilon r}{c_5} > 0, \label{Iaj}
\end{equation}
giving $\epsilon =  c = +1$ and $c_5>0$, which we can thus rename $c_5^2$.

\subsubsection{Final form of the class I solutions}

By implementing the axisymmetry, the no angular deficit and the real number conditions, we obtain the final form of the solutions as
\begin{equation}
l = r \left(2 + \frac{r}{c_5^2}\right),  \label{Ifin1}
\end{equation}
\begin{equation}
\textrm{e}^{\mu} =  \frac{1}{r \left(2  + \frac{r}{c_5^2}\right)}, \label{Ifin2}
\end{equation}
\begin{equation}
k = r \left(2 + \frac{r}{c_5^2}\right) \left[c_k + \frac{2 c_6\left(1 + \frac{r}{c_5^2}\right)}{\sqrt{r \left(2 + \frac{r}{c_5^2}\right)}}\right], \label{Ifin3}
\end{equation}
\begin{equation}
f = c_6^2 - r \left(2 + \frac{r}{c_5^2}\right) \left[c_k + \frac{2 c_6\left(1 + \frac{r}{c_5^2}\right)}{\sqrt{r \left(2 + \frac{r}{c_5^2}\right)}}\right]^2, \label{Ifin4}
\end{equation}
\begin{equation}
D = c_6 \sqrt{r \left(2 + \frac{r}{c_5^2}\right)}, \label{Ifin5}
\end{equation}
\begin{equation}
P_r  =  - \frac{1}{\kappa c_5^2}, \label{Ifin6}
\end{equation}
\begin{eqnarray}
\rho &=& \frac{3}{\kappa c_6^2 c_5^2 r \left(2 + \frac{r}{c_5^2}\right)} \left\{\left[c_5^2 +  2 (1 - c_5^2)r + \left(\frac{1}{c_5^2} - 5 \right) r^2 - \frac{4}{c_5^2} r^3 - \frac{r^4}{c_5^4} \right] \right. \nonumber \\
&\times& \left. \left[c_k + \frac{2 c_6\left(1 + \frac{r}{c_5^2}\right)}{\sqrt{r \left(2 + \frac{r}{c_5^2}\right)}}\right]^2 - c_6^2 c_5^2 \right\}, \label{Ifin7}
\end{eqnarray}
\begin{equation}
v = \frac{1}{c_6}, \label{Ifin8}
\end{equation}
\begin{equation}
\Omega = - \frac{1}{c_6} \left[c_k + \frac{2 c_6\left(1 + \frac{r}{c_5^2}\right)}{\sqrt{r \left(2 + \frac{r}{c_5^2}\right)}}\right], \label{Ifin9}
\end{equation}
\begin{equation}
\dot{V}_1 = \dot{V}^{\alpha}\dot{V}_{\alpha}= 0, \label{Ifin10}
\end{equation}
\begin{equation}
\omega = 0, \label{Ifin11}
\end{equation}
\begin{equation}
\sigma^2 = \frac{1}{r\left(2 + \frac{r}{c_5^2}\right)}. \label{Ifin12}
\end{equation}

\subsubsection{Singularities}

The metric function $\textrm{e}^{\mu}$ diverges on the axis where $r=0$, while $k$, and $l$ vanish at this location. However, since $\rho \stackrel{0}{=} 0$ and $P_r$ is constant, this singularity can be considered as a mere coordinate singularity. Class I solutions are therefore physical singularity-free.

\subsubsection{Lorentzian metric signature} \label{msI}

Since $\textrm{e}^{\mu}$ is obviously positive definite, $f$ must also be positive, which implies, from (\ref{Ifin4}),
\begin{equation}
c_6^2 > r \left(2 + \frac{r}{c_5^2}\right) \left[c_k + \frac{2 c_6\left(1 + \frac{r}{c_5^2}\right)}{\sqrt{r \left(2 + \frac{r}{c_5^2}\right)}}\right]^2. \label{Isign1}
\end{equation}
The right hand side of (\ref{Isign1}) vanishes on the axis. Its behaviour off this region is driven by the respective values of the independent parameters, $c_6$, $c_k$, and $c_5^2$. If it diminishes for increasing $r$, it becomes negative and the inequality is satisfied. If it increases up to equalizing $c_6^2$ for a given value $r_1$ of the radial coordinate defined by
\begin{equation}
c_6^2 = r_1 \left(2 + \frac{r_1}{c_5^2}\right) \left[c_k + \frac{2 c_6\left(1 + \frac{r_1}{c_5^2}\right)}{\sqrt{r_1 \left(2 + \frac{r_1}{c_5^2}\right)}}\right]^2, \label{Isign2}
\end{equation}
this imposes a limit on the radial coordinate, $r_{\Sigma}$, defining the bordering cylinder, which reads
\begin{equation}
r_{\Sigma} <r_1. \label{Isign3}
\end{equation}

In both cases, the metric signature is Lorentzian, and the manifolds are proper GR spacetimes.

\subsubsection{Energy conditions}

In most of the standard applications of GR, the weak energy condition, $\rho>0$, is needed. Here, its expression issued from (\ref{Ifin7}) is a rather complicated formula involving the three parameters defining class I. A detailed analysis of the various possible cases being out of the scope of the present study, we leave it to each particular application.

Owing to (\ref{Ifin6}), $P_r$ is negative. Now, a negative pressure is not specially awkward since this feature is exhibited by a number of physical systems. Moreover, this pressure being constant can be assimilated to a cosmological constant $\Lambda$.

\subsubsection{Junction condition}

The junction condition for a matching of the solutions to a vacuum Lewis-Weyl exterior is recalled in Sec. \ref{junct} as $P_r  \stackrel{\Sigma}{=} 0$. Indeed, (\ref{Ifin6}) implies a constant nonzero pressure all over the spacetime, while, owing to the junction condition, this pressure should vanish on the boundary $\Sigma$. Therefore, this junction condition cannot be verified but we can anyhow consider the negative constant pressure as some cosmological constant.

\subsubsection{Parameters}

Class I is composed of three-parameter solutions. The $c_6$ parameter determines the inverse of the constant velocity $v$ of the fluid, while $c_5^2$ is linked to the magnitude of the shear. The interpretation of $c_k$ is less obvious and we leave it to future work.

\subsection{Class II} \label{cII}

The equation defining this class of solutions reads
\begin{equation}
\mu' = - \frac{(1-h)}{3+h}\frac{l'}{l}, \label{IIa}    
\end{equation}
which, substituted into the Bianchi identity (\ref{Bianchi4r}), gives
\begin{equation}
\frac{h}{1+h} \frac{P_r'}{P_r} - \frac{(3-h)}{2(3+h)} \frac{l'}{l} + \frac{D'}{D} =0, \label{IIc}    
\end{equation}
and, into (\ref{IIIg}), yields
\begin{equation}
\frac{l''}{l} + \frac{l' D'}{l D} - \frac{1}{3+h} \frac{h' l'}{l} - \frac{l'^2}{l^2} +(3+h) \frac{c^2}{l^2}  =0. \label{IId}    
\end{equation}

By inserting (\ref{IIa}) into (\ref{plus1}), we obtain
\begin{equation}
\frac{D''}{D} = \frac{(1+h)}{3+h}\frac{l' D'}{l D} - \frac{l'^2}{4l^2} + \frac{c^2}{l^2}. \label{IIe}    
\end{equation}

We have therefore a set of five simplified independent differential equations, (\ref{IIa})-(\ref{IIe}), plus
the one obtained by the addition of (\ref{G11r}) and (\ref{G22r}), which reads
\begin{equation}
\frac{D''}{D} = \kappa P_r \textrm{e}^{\mu}, \label{IIf}
\end{equation}
for five unknowns, $l$, $\mu$, $D$, $P_r$ and $h$. Once these unknowns are determined, the other metric functions and properties of the fluid follow. This set of equations cannot be solved analytically, even for $h=const.$, unless $h=-1$ which has been already studied as class I. Anyhow, we display this simplified set in view of possible further numerical integration.

\subsection{Class III} \label{cIII}

The defining equation of this class of solutions is
\begin{equation}
\mu'' - \frac{l''}{l} + \frac{\mu' l'}{2l} + \frac{l'^2}{2l^2} - \frac{4 c^2}{l^2} = 0. \label{IIIa}
\end{equation}

These solutions can be integrated along the lines already provided above. We leave this straightforward task to the interested reader. Indeed, the result implies that these solutions exhibit both a negative pressure and a negative energy density, and, therefore verifies neither the weak nor the strong energy condition. Moreover, the only region where the pressure vanishes is at the symmetry axis. This implies that class III spacetimes, are thread-like cylinders, i. e., with no interior, which is a serious shortcoming. Still more serious is the fact that the metric signature is not Lorentzian, at least near the symmetry axis. For all these reasons class III solutions must be ruled out.

\subsection{Class IV} \label{cIV}

The defining equation of this class of solutions is
\begin{equation}
\frac{4c^2}{l^2} - \frac{l'^2}{l^2} = \frac{2\left[(1+h)\mu' + (1-h)\frac{l'}{l}\right] \left(\mu'' - \frac{l''}{l}\right)}{(3+h)\mu' + (1-h)\frac{l'}{l}}, \label{IVa}
\end{equation}
which we insert into (\ref{IIIb}) and obtain
\begin{equation}
\frac{D'}{D} = - \frac{(1-h)\left(\mu'' - \frac{l''}{l}\right)}{(3+h)\mu' + (1-h) \frac{l'}{l}}. \label{IVb}
\end{equation}

Now, (\ref{IIIg}) can be written as
\begin{equation}
\mu'' - \frac{l''}{l} = \frac{4D''}{D} - \left(3 \mu' + \frac{l'}{l}\right) \frac{D'}{D}, \label{IVc}
\end{equation}
which can be substituted into (\ref{IVb}) so as to yield
\begin{equation}
\frac{D''}{D'} = - \frac{h}{1-h}\mu'. \label{IVd}
\end{equation}
This equation can be easily integrated in the case where $h=const.$ Indeed, since only three among the four above equations are independent, adopting such a particular equation of state would suppress any further degree of freedom of the problem. Anyhow, the equations are entangled in a such a manner that the only analytic solution available is for $h=1$ (stiff fluid) and, in this case, the pressure and thus the energy density vanish and we are brought back to a vacuum spacetime.

 The equations determining Class IV have thus been somehow simplified but they still remain analytically unsolved, and, presumably, unsolvable, save numerically.

\section{Conclusions} \label{concl}

The investigations concerning interior spacetimes gravitationally sourced by stationary fluids exhibiting cylindrical symmetry and anisotropic pressure, which have been previously considered for rigidly rotating fluids \cite{C21,C23b,C23c,C23d}, have been extended here to the case of differential rotation. The three configurations where the principal stresses are vanishing in pair are studied for non rigidly rotating irrotational fluids. Irrotationality is the assumption chosen here to simplify, in a physically warranted manner, the problem, thus left with only one degree of freedom.

Having first displayed a general method for obtaining solutions of the GR field equations in the case where the pressure is axially directed, we have applied it to the a particular well-behaved set of solutions which exhibit a large panel of essential physical properties. This set, named class A, can be considered as a remarkable ensemble of solutions, since all the other trials that we have made - more than forty - resulted in fully integrated but ill-behaved solutions, mainly because their energy density was negative. Class A is a set of four-parameter solutions which can be matched to an exterior Lewis-Weyl vacuum and have been shown to satisfy the weak and strong energy conditions. Its hydrodynamical quantities have been calculated and it appears as singularity-free.

As regards the azimuthally directed pressure cases, two different classes have been identified. Class 1, for which a general method of integration has been proposed, is composed of the spacetimes where the ratio $h$ of the pressure over the energy density is a function of the radial coordinate $r$. An expression for this ratio has been exhibited as an explicit analytical function of $r$. Thus the metric and the different quantities representing the physical properties of the system have been provided as functions of $h$, easily convertible in functions of $r$ through $h(r)$.  Class 2 contains the solutions where $h=const.$ It splits into two subclasses, 2i where $h\neq 1$, and 2ii for $h=1$. Provided that the weak energy condition is satisfied, which is generally the case for standard physical applications, class 1 alone allows  a positive pressure, while for subclass 2i the pressure is negative. However, a negative pressure is encountered in various physical systems which allows us to consider this class as a set of proper solutions. As regards subclass 2ii, it displays a negative energy density which is less standard in physics. Anyhow, we have provided it for completeness, and as possibly useful for exotic applications. To reinforce the potential of these solutions to be used as tools designed to represent physical objects encountered in the Universe, they have also been matched on their cylindrical boundary to a Lewis-Weyl exterior vacuum solution.

Contrary to what happens for axially and azimuthally directed pressure, the solutions describing the radial pressure configurations are fully determined by the field equations. Since their manipulation leads to a single equation composed of four vanishing factors corresponding to four differential equations, the last degree of freedom must obligatorily be used to select one or the other among them. We have been therefore confronted to four classes of spacetimes. Class I is composed of fully integrated proper GR spacetimes, whose various mathematical and physical properties have been examined. However, their junction condition to a stationary rotating vacuum exterior cannot be verified as such but one can anyhow consider their negative constant pressure as some kind of cosmological constant. In this case the pressure vanishes, the fluid becomes dust with a cosmological constant and can be matched to the vacuum on the cylinder boundary. Class III has also been fully integrated and it is indeed a solution of the field equations. However, as it is well known, this is not sufficient to determine proper spacetimes and, since the solutions of this class do not exhibit a Lorentzian metric signature, we have been lead to rule them out. Classes II and IV give each, after partial integrations, a set of simplified equations which might be used as seeds for numerical relativity purposes.

Now, these new classes of cylindrically symmetric solutions to field equations of GR sourced by fluids verifying some particular equations of state, together with the ones previously displayed \cite{C21,C23a,C23b,C23c,C23d,C23e}, must be viewed as initiating a step by step discovery of interior spacetimes equiped with more and and more widespread mathematical and physical features, which we hope to pursue in the future.

\appendix

\section{Axisymmetry and angular deficit}\label{secA1}

The axisymmetry condition \cite{S09,G09} has been applied to each of the fully integrated classes of cylindrical solutions for both cases of rigid and differential rotation. In the case of rigid rotation of a radial pressure fluid \cite{C23d}, class Ir has proven compatible with this axisymmetry condition but not with the so-called ``regularity condition'', while, since its Kretschmann scalar does not diverge at the axis, this locus must be considered as a mere coordinate singularity. In the case of differential rotation, we have shown here that present class Id verifies both the axisymmetry and the ``regularity'' conditions.

Now, the ``regularity'' condition method proposed in \cite{M93}, with the aim of avoiding singularities on the axis, consists in checking whether the ratio of the circumference over the radius of an infinitesimally small circle around the axis departs or not from the value $2 \pi$. The circle being the orbit of the spacelike Killing vector $\vec{\xi}$ generating the isometry, this vector must therefore satisfy the so-called ``regularity'' condition \cite{S09}
\begin{equation}
\frac{\partial_{\alpha}\left(\xi^{\mu} \xi_{\mu}\right) \partial ^{\alpha}\left(\xi^{\nu} \xi_{\nu}\right)}{4 \xi^{\lambda}\xi_{\lambda}} \stackrel{0}{=} 1, \label{r0b}
\end{equation}
which, for cylindrical symmetry as described here, becomes (\ref{Iad}). 

Moreover, this ``regularity'' condition is obtained by a normalisation of the $\phi$ coordinate provided that elementary flatness is assumed in the vicinity of the axis. However, it has been shown in \cite{L94} that elementary flatness by no means guarantees regularity. Moreover, it has been claimed in \cite{W96} that, regularity being defined in Cartesian type coordinates, a violation of the regularity condition might be linked to some issue with the use of polar type coordinates. The compatibility of polar type coordinates with the study of features of the axis has also been discussed in \cite{C00}, while the presence of some matter on the axis has been referred to in \cite{P96} to deal with the regularity problem. 

Since this so-called ``regularity'' condition is designed to prevent an angular deficit possibly due to the presence of conical singularities on the axis, we have proposed, in \cite{C23e}, to consider henceforth this condition as an angular deficit avoidance condition, named ''non-angular deficit condition''.

Thus, the behaviour of ``rigid'' class I reinforce the above statement that an angular deficit in the vicinity of $r=0$ does not obligatorily imply a singular axis. Moreover, by comparing the behaviour of the rigid and differentially rotating configurations, we are led to postulate that passing from rigid to differential rotation of the gravitational source affects the properties exhibited by the spacetimes in the vicinity of the axis. Indeed, when the pressure is axially or azimuthally directed, the rigidly rotating solutions are devoid of any angular deficit near the axis, while they exhibit such a feature when the rotation is differential. It is the reverse when a radial pressure is involved.

\end{document}